\documentclass[aps,prl,reprint,superscriptaddress]{revtex4-2}

\usepackage{amsmath} 
\usepackage{graphicx}
\usepackage{dcolumn}
\usepackage{bm}
\usepackage[utf8]{inputenc}
\usepackage[T1]{fontenc}
\usepackage{physics} 
\usepackage{color,soul}
\usepackage{comment}
\usepackage[colorlinks = true, linkcolor = blue, urlcolor  = blue, citecolor = blue, anchorcolor = blue]{hyperref}
\usepackage{moreverb} 
\usepackage{mathtools}
\usepackage{cancel}
\usepackage[capitalise]{cleveref}
\newcommand{\papertitle}{Transonic and supershear crack propagation driven by geometric nonlinearities}

\newcommand{\crackspeed}{c_\mathrm{f}}
\newcommand{\Rwavespeed}{c_\mathrm{R}}
\newcommand{\Swavespeed}{c_\mathrm{s}}

\newcommand{\fractureenergy}{\Gamma}
\newcommand{\energyreleaserate}{G}
\newcommand{\staticenergyreleaserate}{\energyreleaserate_0}
\newcommand{\universalfunction}{g(\crackspeed)}

\newcommand{\strain}{E}
\newcommand{\straintensor}{\strain_{ij}}
\newcommand{\infinitesimalstrain}{\varepsilon}
\newcommand{\gradientofdisplacement}{e}
\newcommand{\infinitesimalstraintensor}[1]{\infinitesimalstrain_{#1}}
\newcommand{\largestrainswitch}{\alpha}
\newcommand{\stresstensor}{\sigma_{ij}}
\newcommand{\cauchy}[1]{\tau_{#1}}
\newcommand{\pkI}[1]{S_{#1}}
\newcommand{\pkII}[1]{\sigma_{#1}}
\newcommand{\greenstrain}[1]{E_{#1}}

\newcommand{\acceleration}{\Ddot{u}}

\newcommand{\dV}[1]{\mathrm{d}#1}
\newcommand{\jump}[1]{[\![#1]\!]}

\newcommand{\density}{\rho}
\newcommand{\shearmodulus}{\mu}
\newcommand{\poissonratio}{\nu}

\newcommand{\remoteload}{\sigma_{\infty}}

\newcommand{\imposedstretch}{\lambda}

\newcommand{\crackhalflength}{a}
\newcommand{\griffithlength}{L_\mathrm{G}}
\newcommand{\crackopening}{\delta}
\newcommand{\criticalopening}{\delta_\mathrm{c}}

\newcommand{\ie}{{\textit{i.e.}}}
\newcommand{\currenttime}{t+ \Delta t}
\newcommand{\currentconfig}{\prescript{\currenttime}{}{\Omega} }
\newcommand{\referenceconfig}{\prescript{0}{}{\Omega} }
\newcommand{\avg}[1]{\langle #1 \rangle }

\begin{document}

\title{\papertitle}
\author{Mohit Pundir}
\affiliation{Institute for Building Materials, ETH Zurich, Switzerland}

\author{Mokhtar Adda-Bedia}
\affiliation{Laboratoire de Physique, CNRS, ENS de Lyon, Universit\'e de Lyon, 69342 Lyon, France}

\author{David S. Kammer}
\email{dkammer@ethz.ch}
\affiliation{Institute for Building Materials, ETH Zurich, Switzerland}
\date{\today}

\begin{abstract} 
Linear elastic fracture mechanics theory predicts that the speed of crack growth is limited by the Rayleigh wave speed. Although many experimental observations and numerical simulations have supported this prediction, some exceptions have raised questions about its validity. The underlying reasons for these discrepancies and the precise limiting speed of dynamic cracks remain unknown. Here, we demonstrate that tensile (mode~I) cracks can exceed the Rayleigh wave speed and propagate at supershear speeds. We show that taking into account geometric non-linearities, inherent in most materials, is sufficient to enable such propagation modes. These geometric non-linearities modify the crack-tip singularity, resulting in different crack-tip opening displacements, cohesive zone behavior, and energy flows towards the crack tip.
\end{abstract}

\maketitle 

The speed at which cracks propagate is a fundamental characteristic that has implications in various fields such as material design~\cite{baumberger_solvent_2006}, earthquake mechanics~\cite{bao_global_2022}, and even the phenomenon of popping balloons~\cite{moulinet_popping_2015}. Linear Elastic Fracture Mechanics (LEFM)~\cite{freund_dynamic_1998} plays a crucial role in predicting crack speed $\crackspeed$ by establishing an energy balance between the energy release rate, which drives crack growth, and the fracture energy ($\fractureenergy$), which resists it. This framework, which assumes that $\fractureenergy$ is dissipated solely at the crack tip, predicts that the material Rayleigh wave speed $\Rwavespeed$ serves as a limiting speed for crack propagation. This prediction has been experimentally confirmed~\cite{goldman_acquisition_2010}. Crack growth occurring at speeds between $\Rwavespeed$ and the shear wave speed $\Swavespeed$ is considered physically inadmissible, as it would generate energy rather than dissipate it. Nevertheless, LEFM predicts that cracks can propagate at supershear speeds $\crackspeed > \Swavespeed$ if one assumes that dissipation occurs within a spatially extended zone around the crack tip~\cite{freund_dynamic_1998,broberg1999cracks}. However, the specific conditions that allow for supershear propagation of, particularly, opening (mode~I) cracks and the processes involved in the transition through the forbidden speed range remain largely unknown.

Supershear crack growth is predominantly observed in cracks under shear (mode~II) loading conditions, as described theoretically~\cite{burridge_admissible_1973,kammer_equation_2018} and widely supported by numerical simulations~\cite{andrews_rupture_1976,abraham_how_2000,kammer_equation_2018}, experimental studies~\cite{rosakis_cracks_1999,xia_laboratory_2004,ben-david_dynamics_2010,passelegue_sub-rayleigh_2013,svetlizky_properties_2016}, and natural observations~\cite{bao_global_2022,bouchon_observation_2003,walker_illuminating_2009,dunham_evidence_2004,wang_fast_2016}. Supershear propagation is generally associated with high-stress states~\cite{andrews_rupture_1976,kammer_equation_2018}. In contrast, supershear propagation in cracks under mode~I loading conditions is relatively rare. Molecular dynamics (MD) simulations~\cite{buehler_hyperelasticity_2003} and lattice models~\cite{guozden_supersonic_2010,slepyan1981crack,marder_shock-wave_2005,marder_supersonic_2006} have shown instances of supershear crack speeds, while experimental observations have been reported for rubber-like materials~\cite{petersan_cracks_2004,moulinet_popping_2015,mai_crack-tip_2020}, hydrogels~\cite{wang_tensile_2023} and structural materials where the loading is applied directly at the crack tip by some extreme conditions~\cite{winkler_crack_1970}. The presence of some type of non-linearity, extending beyond the limits of LEFM, is a recurring feature in both simulations and experiments, indicating its potential contribution to enabling supershear growth in tensile cracks. However, the specific type of material non-linearity required for this phenomenon, as well as its generality across different materials, remains unknown.

Here, we investigate the minimal requirements for the transition to supershear propagation of tensile cracks using numerical simulations. Our simulations reveal that the presence of geometric non-linearities alone is the primary factor driving supershear crack growth resulting from a continuous acceleration through the transonic speed range. Since such non-linearities are generally present in materials, these findings demonstrate that supershear propagation is an inherent characteristic in dynamic crack problems, independent of the specific material constitutive laws. 

\begin{figure}[htb]
\centering
\includegraphics[width=0.98\columnwidth]{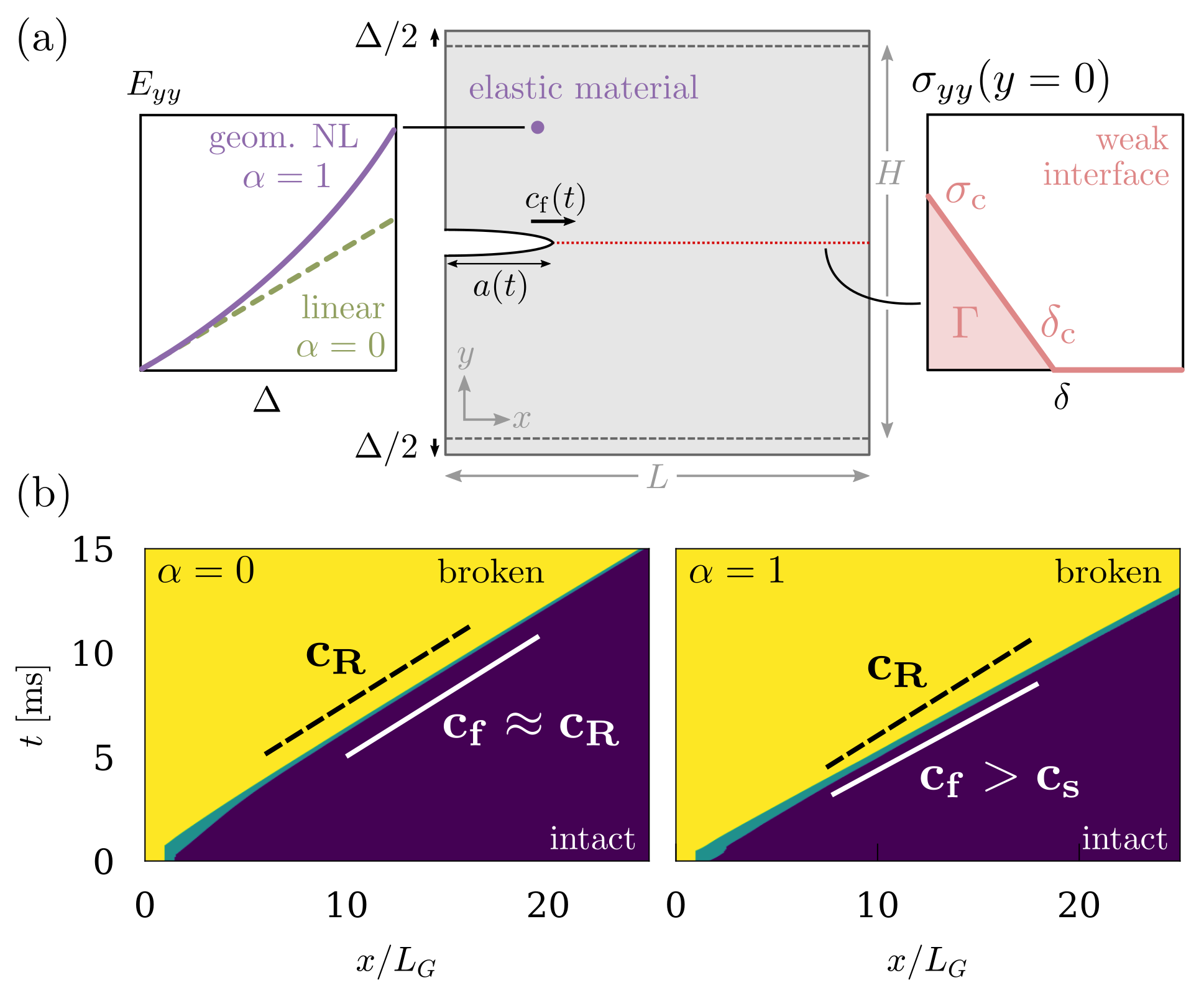}
\caption{Model setup and illustrative examples. (a) 2D model configuration with an elastic material and a weak cohesive interface. (b) Temporal evolution of interface for simulations at an imposed stretch $\imposedstretch = 1.125$ with (left) linear elastic material ($\alpha=0$) and (right) geometric non-linear elastic material ($\alpha=1$). Blue indicates intact material, turquoise the cohesive zone area, and yellow the broken interface. The crack speed $\crackspeed$ is indicated by white lines, and the Rayleigh wave speed $\Rwavespeed$ by a black dashed line.}
\label{fig:setup}
\end{figure}

We consider the most generic and simple model without introducing a non-linear material constitutive law or any additional material parameter. The material deformation is described by a two-dimensional plane-strain tensor $\straintensor$, defined as
\begin{equation}
    \straintensor = \frac{1}{2} \left( \frac{\partial u_i}{\partial x_j} +  \frac{\partial u_j}{\partial x_i} + \largestrainswitch \frac{\partial u_k}{\partial x_j} \frac{\partial u_k}{\partial x_i} \right)~,
    \label{eq:strain}
\end{equation}
where $u_i$ and $x_i$ are the $i$-th displacement and coordinate component $(i\equiv x,y)$, respectively. To easily switch between linear and geometrically non-linear (GNL) cases, we introduce the factor $\largestrainswitch \in \{0,1\}$. Therefore, for $\largestrainswitch=1$, $\straintensor$ corresponds to the Green-Lagrangian strain tensor, while for $\largestrainswitch=0$, it is its linear approximation, the infinitesimal strain tensor $\infinitesimalstraintensor{ij}$ (see \cref{fig:setup}a).
In our model, we employ a linear elastic constitutive law, as described by the shear modulus $\shearmodulus$ and Poisson's ratio $\poissonratio$, to relate the two-dimensional $2^{\mathrm{nd}}$-Piola-Kirchhoff stress tensor $\stresstensor$ to $\straintensor$, through
\begin{equation}
    \stresstensor = 2\shearmodulus \left(\straintensor + \frac{\poissonratio}{1-2\poissonratio} \delta_{ij} \strain_{kk} \right)~,
    \label{eq:constitutivelaw}
\end{equation}
where $\delta_{ij}$ is the Kronecker delta. Notice that, as sketched in \cref{fig:setup}a, the GNL model ($\alpha=1$) induces a strain-enhancing effect with respect to the linear case ($\alpha=0$). The advantage of this model, compared to one with a non-linear constitutive law (\textit{e.g.}, neo-Hookean), is that it allows isolation of the effect of non-linearity on the crack propagation. In the following, we choose $\shearmodulus=39.2~\mathrm{kPa}$ and $\poissonratio=0.35$, and then solve the problem for the conservation of linear momentum (full details provided in~\cite{SM}). 

Fracture of the material is modeled using a cohesive approach, where cohesive tractions across the crack plane represent the progressive failure of the material. In our model, see \cref{fig:setup}a, we adopt a linear cohesive law with $\sigma_c = 20~\mathrm{kPa}$ and $\Gamma =15~\mathrm{J/m^2}$, which gives a critical opening distance of $\criticalopening = 1.5~\mathrm{mm}$ (for details, see~\cite{SM}). This cohesive approach allows for the representation of a cohesive zone that captures localized spatially distributed dissipation, providing an approximation of the process zone observed in natural fractures. We use standard numerical techniques, detailed in~\cite{SM}, to accurately simulate crack growth.

We examine the behavior of fracture growth in a two-dimensional plane-strain system of height $H=102.6~\mathrm{mm}$ and length $L=154~\mathrm{mm}$ mimicking the most common experimental configuration (see \cref{fig:setup}a). The dimensions are chosen sufficiently large to avoid any wave reflections that could affect the results. 
We apply a uniform and constant remote displacement $\Delta \gg \criticalopening$, which results in a uniform stretch $\imposedstretch=1+(\Delta/H)$ on the entire sample. We initiate crack growth at time $t = 0$ by artificially introducing a seed crack that slightly exceeds Griffith's critical length for plane-strain conditions given by $\griffithlength  =2\shearmodulus \Gamma/\pi (1-\nu) \remoteload^2 = 5.1~\mathrm{mm}$ where $\remoteload$ is the applied stress induced by the imposed remote displacement $\Delta$. The growth of the crack is confined to a (weak) plane perpendicular to the imposed stretch and aligned with the seed crack, restricting its propagation to a straight path $y=0$ (as illustrated in \cref{fig:setup}a). This constraint effectively prevents crack branching instabilities, commonly observed~\cite{washabaugh_reconciliation_1994,sharon_confirming_1999}, imitates the grooves used in experiments~\cite{washabaugh_reconciliation_1994,wang_tensile_2023}, and enables a thorough exploration of crack speeds across the full range.

First, we consider the linear elastic case ($\alpha = 0$). Immediately after the seed crack is introduced, it becomes unstable, accelerates, and propagates through the entire interface (\cref{fig:setup}b-left). The crack-tip position, as defined by the transition from intact material to the cohesive zone (see \cref{fig:setup}b-left), moves through the interface, leading to a growing crack length $\crackhalflength(t)$. We observe that the crack speed, as computed by $\crackspeed = \mathrm{d} \crackhalflength / \mathrm{d} t$, approaches $\Rwavespeed$ (see \cref{fig:setup}b-left) but does not exceed it respecting the limiting speed given by LEFM. Considering the exact same model with the sole difference of including geometric nonlinearities ($\alpha = 1$), we observe a different crack propagation (\cref{fig:setup}b-right). In this case, the crack continuously surpasses both $\Rwavespeed$ and $\Swavespeed$, and propagates at supershear speeds. These results reveal an unknown mechanism for supershear propagation, which is simply due to geometric non-linearities.

\begin{figure}[htb]
\centering
\includegraphics[width=0.98\columnwidth]{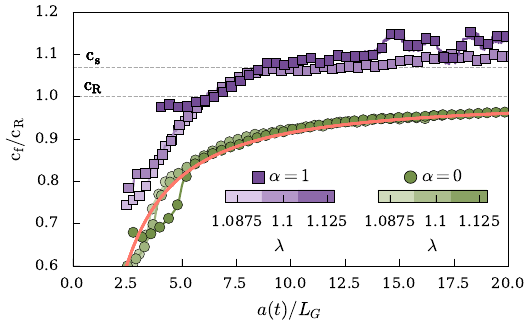}
\caption{Crack-tip dynamics for three different values of stretches $\lambda=1.0875, 1.1$ and $1.125$. The crack dynamics are shown in green shades for $\alpha=0$, and purple shades for $\alpha=1$, respectively. The red solid line is the LEFM crack-tip equation of motion.}
\label{fig:crack-dynamics-different-stretches}
\end{figure}

For a quantitative evaluation of the different crack growth behavior, we consider the instantaneous crack-tip dynamics, as shown in \cref{fig:crack-dynamics-different-stretches}. For uniform systems, LEFM predicts that $\crackspeed / \Rwavespeed \approx 1 - \griffithlength/\crackhalflength$ (details provided in~\cite{SM}), which shows that a LEFM-governed crack remains sub-Rayleigh -- even if it gets infinitely long. We observe that the simulation with a linear elastic material ($\alpha=0$) agrees quantitatively well with the LEFM prediction (\cref{fig:crack-dynamics-different-stretches}). Specifically, it asymptotically approaches $\Rwavespeed$, satisfying the theoretical limit by remaining at sub-Rayleigh speeds for all crack lengths. In contrast, the simulations with geometric non-linearities ($\alpha = 1$) do not follow the LEFM prediction and exceed the limiting speed $\Rwavespeed$. 

These results reveal a few important mechanisms for crack dynamics of geometrically non-linear materials. First, simulations at different stretch levels are superimposed when normalized by Griffith's length (see \cref{fig:crack-dynamics-different-stretches}), which suggests that there is a crack-tip equation of motion for geometrically non-linear materials. Second, the crack speeds in the GNL case are consistently and significantly above the LEFM prediction even in the sub-Rayleigh regime (\ie{} $\crackspeed < \Rwavespeed$), indicating that the LEFM energy balance is fundamentally changed. Third, the crack accelerates simply through the transonic speed range $\left[\Rwavespeed,\Swavespeed\right]$ to reach supershear speeds ($\crackspeed > \Swavespeed$). This is fundamentally different from the subRayleigh-to-supershear transition observed in shear cracks, in which a secondary crack ahead of the main crack is required to allow for a crack speed jump (\ie{} discontinuity) across the forbidden transonic speed range~\cite{andrews_rupture_1976,xia_laboratory_2004,svetlizky_properties_2016}. 

\begin{figure}[htb]
\centering
\includegraphics[width=0.98\columnwidth]{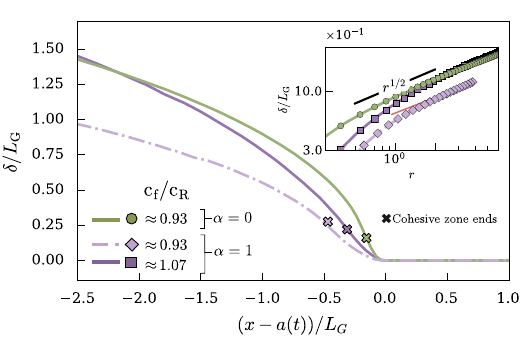}
\caption{Crack-tip opening displacement,  $\delta$,  for simulations at $\imposedstretch=1.1$ with small strain $\alpha=0$ (shown in green) and geometric non-linear strain $\alpha=1$ (shown in purples), respectively. For $\alpha=1$, the light purple corresponds to the subRayleigh  speed, $ \crackspeed \approx 0.93\,\Rwavespeed$, and the dark purple corresponds to subsonic speed, $\crackspeed \approx 1.07\,\Rwavespeed$. The cross marks indicate the ends of the cohesive zone. (inset) A log-log plot of $\delta$ as a function of $r=(a(t)-x)/L_G$, the distance from the crack tip. The black and red lines indicate $r^{1/2}$ scaling.}
\label{fig:k-tip-morphology}
\end{figure}

Next, we focus on the crack-tip opening displacement to determine the mechanisms allowing for propagation through the transonic regime. For the linear case ($\alpha=0$), see \cref{fig:k-tip-morphology}, the crack opening $\crackopening$ follows a square-root behavior outside the cohesive zone, which is, as expected, consistent with LEFM. The GNL material, however, presents a different behavior. Close to the crack-tip (but outside the cohesive zone), the exponent increases (see inset in \cref{fig:k-tip-morphology}). This effect becomes even stronger when the crack surpasses $\Rwavespeed$ (see \cref{fig:k-tip-morphology}). These results suggest that the square-root singular behavior of the strain and stress fields is not relevant when GNL effects are considered. This calls into question the foundations of brittle fracture mechanics, which are based mainly on the consequences of the square root behavior for energy budgeting and, thus, for the crack equation of motion. Such an observation may waver one of the fundamental corollaries of LEFM: the maximum speed allowed by the rate of energy flow toward the crack tip.

\begin{figure*}[t]
\centering
\includegraphics[width=1.98\columnwidth]{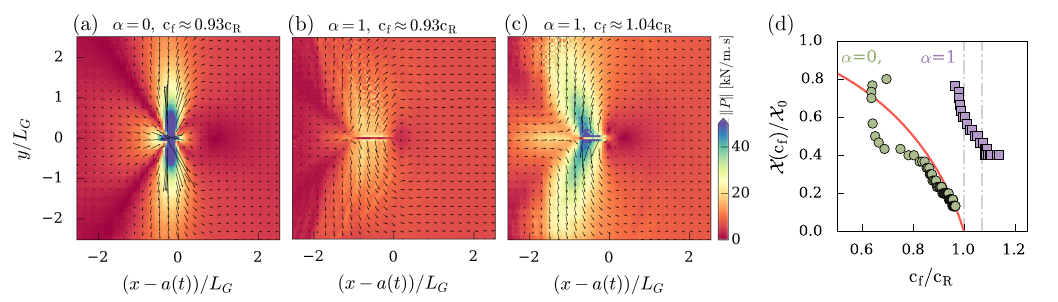}
\caption{Snapshots of energy flow density, represented as the magnitude of Poynting vector $P_{j}$ (see \cite{SM} for more details). All the snapshots are for an imposed stretch $\imposedstretch = 1.125$ (a) $\alpha=0$  at subRayleigh speed $\crackspeed\approx 0.93 \Rwavespeed$ (b) $\alpha=1$ at subRayleigh speed, $\crackspeed\approx 0.93 \Rwavespeed$ (c) $\alpha=1$ at subsonic speed, $\crackspeed\approx 1.04  \Rwavespeed$. (d) Evolution of cohesive zone size $\mathcal{X}(\crackspeed)$ for $\imposedstretch=1.125$ with small strain $\alpha=0$ and geometric non-linear strain $\alpha=1$, respectively. The cohesive zone size is normalized by static cohesive zone size $\mathcal{X}_0$. The red solid line shows the analytical solution for cohesive zone size from LEFM.}
\label{fig:cohesive-zone-size}
\end{figure*}

To visualize further the modified near-tip fields and the associated energy flow to the crack tip, we compute the Poynting vector~\cite{freund_dynamic_1998,buehler_hyperelasticity_2003} (details in~\cite{SM}). For the linear material, the Poynting vector field takes the ordinary shape and values (see \cref{fig:cohesive-zone-size}a). At the same sub-Rayleigh crack speed, the GNL material presents a significantly different pattern (see \cref{fig:cohesive-zone-size}b). While the magnitude of the Poynting vector is lower in the vicinity of the crack tip (note the absence of blue color), it is somewhat increased further away (see brighter red at $|x| > \crackhalflength)$. Further changes occur at transonic crack speeds (see \cref{fig:cohesive-zone-size}c), where the magnitude of the Poynting vector increases but remains below the values observed in the linear material, and the lobes are inclined to the back of the crack, which are forerunners of the Mach cone in the supershear propagation regime~\cite{buehler_hyperelasticity_2003}. These modifications to the near-tip crack fields confirm that the crack dynamics changed due to the GNL material behavior and point to a totally different energy budgeting even in the subshear propagation regime.

The modified energy flux to the crack tip also causes changes to the local dissipation, which manifests itself in the properties of the cohesive zone. From the near-tip Poynting vector fields (see \cref{fig:cohesive-zone-size}a-c), we observe an increase in the cohesive-zone size $\mathcal{X}(\crackspeed)$ for the GNL case. Quantitatively, the cohesive zone size for the linear material follows, after some initial perturbations from the nucleation, the LEFM prediction~\cite{freund_dynamic_1998} with a Lorentz contraction from its static size $\mathcal{X}_0$ to zero towards $\Rwavespeed$ (see \cref{fig:cohesive-zone-size}d). In contrast, the cohesive zone in the GNL material is considerably larger and appears to be relatively constant at $\mathcal{X} \approx 0.4\,\mathcal{X}_0$ while the crack traverses the transonic regime and reaches supershear propagation. This suggests that kinetic energy and bulk wave interference responsible for the Lorentz contraction in LEFM yield completely different results in materials where GNL effects are dominant. The fact that $\mathcal{X}$ is finite in the transonic regime indicates that the Lorentz contraction is superseded by a different mechanism that could be related to either the cohesive zone response or the non-square root singular behavior near the crack tip.

The present study confirms previous numerical simulations related to the limiting speed of propagating cracks~\cite{buehler_hyperelasticity_2003,guozden_supersonic_2010,slepyan1981crack,marder_shock-wave_2005,marder_supersonic_2006}. In MD simulations, where hyperelasticity was assumed, supershear propagation was related to an increased local wave speed in the highly stretched region at the crack tip compared to the far-field wave speed. In simulations using a lattice model, a completely different branch of supershear solutions for propagating cracks was found for stretches above a critical value. However, the novelty of the present work is to show that supershear propagation is possible within a continuum elastic framework. We show that GNL is the principal ingredient to ``break'' the barrier of Rayleigh wave speed for tensile crack propagation and that there is no forbidden interval velocity. Tensile cracks accelerate smoothly from the sub-Rayleigh regime to the supershear regime. These observations are robust as they do not depend on the specific choice of material parameter values (see \cite{SM}). Moreover, they agree with experiments on crack propagation in hydrogels~\cite{wang_tensile_2023} and rubber-like materials~\cite{petersan_cracks_2004,mai_crack-tip_2020}. The common thread between our model and these experiments is that the materials considered exhibit a nonlinear elastic response, for which LEFM is possibly not an adequate framework. Note that in the case of hydrogels, experiments show that supershear rupture is enabled above an applied stretch level~\cite{wang_tensile_2023}.  We believe that the existence of such a critical stretch is caused by the velocity dependence of the fracture energy of the hydrogel. Our model is consistent with a constant $\fractureenergy$ that allows the crack to always be in an accelerating phase (both in LEFM and certainly in GNL frameworks). For GNL simulations, our system size and other crack nucleation aspects prevent us from determining the terminal supershear speed of the crack, which is beyond the scope of this study.

Let us conclude with a discussion on possible directions for future work. The results for the crack-tip opening displacement, energy release rate, and cohesive zone size point toward the conclusion that the elastic field distribution and energy budgeting in the vicinity of the crack tip of a GNL material exhibit completely different behaviors than that of a linear elastic material. Recent attempts to uncover the underlying mechanisms tackled such problems perturbatively~\cite{livne_near-tip_2010}; however, our results lean toward a non-perturbative effect. We believe that the methods developed for static cracks in nonlinear materials~\cite{long_crack_2015} should be generalized to the dynamic problem. Indeed, important questions arise related to a material's non-linear response which produces either strain stiffening or strain softening at high stretches. How does it affect our findings and is any non-linearity capable of allowing supershear cracks? For example, it is believed that dynamic crack growth in a purely neo-Hookean material follows LEFM solutions~\cite{livne_near-tip_2010}. How does this reconcile with the fact that the non-linearity used in our model is present in any material? Why is it believed that supershear crack propagation is not possible in engineering brittle materials such as glass? For this, other instabilities may occur at lower speeds (such as microbranching, oscillatory instabilities, and instabilities along the crack front) prevent cracks from reaching transonic and supershear speeds. Moreover, the brittleness of such materials does not allow them to store enough potential energy prior to crack propagation. This would explain why supershear cracks are commonly associated with direct extreme loading on the crack surface~\cite{winkler_crack_1970,rosakis_cracks_1999}. 

Finally, the classical LEFM theory developed over the last century~\cite{freund_dynamic_1998,broberg1999cracks} has been a strong backbone for our understanding of material fracture. Here, we demonstrate its breakdown by showing that naturally existing geometric non-linear strain causes supershear crack propagation. This study sets a starting stone for the development of a new framework within the mechanics of continuous media: nonlinear elastic fracture mechanics (NLEFM).

\begin{acknowledgments}
DSK and MP acknowledge support from the Swiss National Science Foundation under the SNSF starting grant (TMSGI2\_211655).
We thank J. Fineberg and M. Wang for sharing their experimental results~\cite{wang_tensile_2023} prior to publication and for fruitful discussions.
\end{acknowledgments}


%

\iftrue
\clearpage
\appendix
\section{\textsc{Supplemental Material}}
\section{Total Lagrangian formulation for geometrical non-linearties}
Let $\cauchy{ij}$ denote the true (Cauchy) stress tensor. The equation of motion in the absence of body forces reads:
\begin{equation} \label{eq:cauchy-eom}
\cauchy{ij, j} - \prescript{\currenttime}{}{\density}\acceleration_i = 0 \quad \text{on}~ \currentconfig
\end{equation}
where 
$\currentconfig$ is the body domain in the deformed (actual) configuration at time $\currenttime$,
$\prescript{\currenttime}{}{\density}$ is the material density in that configuration, and the differentiation is with respect to the deformed configuration. 
The weak form of \cref{eq:cauchy-eom} reads 
\begin{equation}
\begin{split}
  \int_{\currentconfig}  \delta u_i \cauchy{ij, j} \mathrm{d}\Omega~ - \int_{\currentconfig} \delta u_i \prescript{\currenttime}{}{\density}\acceleration_i \mathrm{d}\Omega = 0 ~.
  \end{split}    
\end{equation}
Applying integration by parts and divergence theorem, while omitting the boundary terms, we find
\begin{equation}\label{eq:weak-form-cauchy}    
  \int_{\currentconfig}  \delta\gradientofdisplacement_{ij} \cauchy{ij}~ \mathrm{d}\Omega  ~+\int_{\currentconfig} \delta u_i\prescript{\currenttime}{}{\density}\acceleration_i ~\mathrm{d}\Omega = 0 ~,
\end{equation}
where $\gradientofdisplacement_{ij}$ is the linear strain tensor in the deformed configuration.

Since a body can undergo large displacements and rotations, the above relation cannot be solved directly. It is solved incrementally where the solution at $\currenttime$ is approximated from already known equilibrium configurations. Thus, it is more suitable to employ the \textit{Total Lagrangian} (T.L.) formulation where the known equilibrium configuration at $t=0$ is considered, and all the kinematic variables are thus computed relative to this initial configuration.  The T.L. formulation uses the $2^{\mathrm{nd}}$-Piola-Kirchhoff stress measure $\pkII{ij}$, and Green-Lagrange strain measure $\greenstrain{ij}$ to formulate \cref{eq:weak-form-cauchy} with respect to the initial configuration at time $t = 0$. This yields
\begin{equation}\label{eq:weak-form}
    \int_{\referenceconfig} \delta\prescript{\currenttime}{}{ \greenstrain{ij}} \prescript{\currenttime}{}{\pkII{ij}~\mathrm{d}\Omega}~ + \int_{\referenceconfig} \delta \prescript{\currenttime}{}{u_i}\prescript{0}{}{\density}\prescript{\currenttime}{}{\acceleration_i} ~\mathrm{d}\Omega = 0
\end{equation}
where for a geometrical non-linear case
\begin{align}
\delta \prescript{\currenttime}{}{\greenstrain{ij}}&=\delta\dfrac{1}{2}(\prescript{\currenttime}{}{u_{i,j}} + \prescript{\currenttime}{}{u_{j,i}} + \prescript{ \currenttime}{}{u_{k,i}}\prescript{\currenttime}{}{u_{k, j}}) ~.
\end{align}
Notice that the differentiation is with respect to the initial configuration.
A complete derivation of \cref{eq:weak-form} from \cref{eq:weak-form-cauchy} can be found in~\cite{bathe_finite_1996}. 

The stress and strain at $\currenttime$ is expressed as a sum of stress and strain at a previous time $t$ and their incremental change at $\Delta t$, \ie{}
\begin{align}
\prescript{\currenttime}{}{\pkII{ij}} &= \prescript{t}{}{\pkII{ij}} + \Delta \pkII{ij}   \\
\prescript{\currenttime}{}{\greenstrain{ij}} &= \prescript{t}{}{\greenstrain{ij}} + \Delta \greenstrain{ij}; ~ \Delta \greenstrain{ij} = \Delta \infinitesimalstrain_{ij} + \Delta\eta_{ij}\label{eq:strain-decomposition}
 \end{align}
 where $\Delta \infinitesimalstrain_{ij}$, $\Delta \eta_{ij}$ is the linear and non-linear incremental strains referred to the configuration at $t = 0$ (for details, see~\cite{bathe_finite_1996}). The stress-strain relationship for a linear-elastic material is given by \cref{eq:constitutivelaw}. 
From \cref{eq:strain-decomposition}, we can say that $\delta \prescript{\currenttime}{}{\greenstrain{ij}} = \delta \Delta \greenstrain{ij}$ and consequently,  linearizing $\delta \Delta\greenstrain{ij}$,  \ie{} assuming $\delta \Delta \greenstrain{ij} = \delta \Delta \infinitesimalstrain_{ij}$, the equation of motion is given as 
\begin{equation*}
    \int_{\referenceconfig} \delta \Delta\infinitesimalstrain_{ij}\prescript{\currenttime}{}{\pkII{ij}~\mathrm{d}\Omega} ~ + \int_{\referenceconfig} \delta \prescript{\currenttime}{}{u_i}\prescript{0}{}{\density}\prescript{\currenttime}{}{\acceleration_i} ~\mathrm{d}\Omega = 0 ~.
\end{equation*}
We employ an explicit time integration (central difference method) along with a predictor-corrector scheme to solve the above equation of motion.
 \par
\noindent
The predictor stage is given as :
\begin{align*}
\prescript{\currenttime}{}{u_i} &= \prescript{t}{}{u_i} + \Delta t \prescript{t}{}{\dot{u}_{i}} + \dfrac{\Delta t^2}{2}\prescript{t}{}{\acceleration_i} \\
\prescript{\currenttime}{}{\dot{u}_i} &= \prescript{t}{}{\dot {u}_i} + \Delta t\prescript{t}{}{\acceleration_i} \\
\prescript{\currenttime}{}{\acceleration_i} &= \prescript{t}{}{\acceleration_i}
\end{align*}
Solving the equation of motion for incremental acceleration $\Delta\acceleration_i$:
\begin{equation}\label{eq:equation-motion}
 \begin{split}
    \int_{\referenceconfig} \delta \Delta u_i \prescript{0}{}{\rho} \Delta \acceleration_i~\mathrm{d}\Omega = - \int_{\referenceconfig} \delta\Delta \infinitesimalstrain_{ij}\prescript{\currenttime}{}{\pkII{ij}~\mathrm{d}\Omega} \\ - \int_{\referenceconfig} \delta \Delta u_i \prescript{0}{}{\density} \prescript{t}{}{\acceleration_i}~\mathrm{d}\Omega
    \end{split}
\end{equation}
The corrector stage is given as :
\begin{align*}
\prescript{\currenttime}{}{\acceleration_i} &= \prescript{t}{}{\acceleration_i} + \Delta \acceleration_i \\
\prescript{\currenttime}{}{\dot{u}_i} &= \prescript{\currenttime}{}{\dot {u}_i} + \Delta t\Delta\acceleration_i \\
    \prescript{\currenttime}{}{u_i} &= \prescript{\currenttime}{}{u_i} 
\end{align*}
We employ the Finite Element method to solve the above equations. To ensure numerical stability, we chose the incremental time step $\Delta t$ based on the Courant–Friedrichs–Lewy condition: $\Delta t  = 0.05\times h_\mathrm{min}\sqrt{\prescript{0}{}{\density}(1-\nu^2)/2\mu(1+\nu)}$, where $h_\mathrm{min}$ is the minimum element size, $\mu$ is the shear modulus of the material, and $\nu$ is its Poisson’s ratio.




\section{Implementation of cohesive elements using \textit{discontinuous Galerkin} method} 
We adopt a \textit{discontinuous Galerkin} finite element method for cohesive zone modelling~\cite{hansbo_finite_2004, hansbo_discontinuous_2015}. We apply Nitsche's method to tie the meshes together, \ie{} to weakly enforce the continuity of displacements across the internal elemental interfaces~\cite{ten_eyck_discontinuous_2006, hansbo_discontinuous_2015}. To this end, additional terms are added to the second term in \cref{eq:equation-motion}. For simplicity, we have dropped the superscript $\currenttime$ and subscript $0$. The replaced expression is given as 
\begin{equation}\label{eq:prior-facture}
\begin{split}
\int_{\Omega} \delta \infinitesimalstrain_{ij}\pkII{ij} \dV{\Omega} &~ - \int_{\Gamma} \left( \jump{v}_{i} n^{+}_{j} \right)\avg{\pkII{ij}} \dV{\Gamma}~ - \\ &\int_{\Gamma} \left( \jump{u}_{i} n^{+}_{j}\right)  \avg{\pkII{ij}(v)} \dV{\Gamma} + \vartheta\int_{\Gamma}  \jump{v}_{i}\jump{u}_{i}\dV{\Gamma} 
 \end{split}
\end{equation}
where $\jump{x} = x^{+} - x^{-}$ represents the jump across the interface ($\Gamma^{+}, \Gamma^{-}$), $\avg{x} = \frac{1}{2}(x^{+} + x^{-})$ represents the average across an interface and $n_j^{+}$ is the outward normal from the interface $\Gamma^{+}$. The second term in the above equation makes the formulation consistent. The third term is added to make the formulation symmetric, which improves convergence, and the fourth term enforces the displacement continuity. The  stabilization parameter $\vartheta$ that ensures that the formulation remains positive-definite. For a linear elastic material, $\vartheta$ depends on the material parameters and the element size $h$, \ie{}  $ \vartheta = \theta \mu(1 + 2\nu/(1-2\nu) )/h$ where $\mu$ is shear modulus, $\nu$ is Poisson's ratio and $\theta$ an arbitrarily chosen positive number~\cite{liu_three-dimensional_2009}. \Cref{eq:prior-facture} is valid only prior to fracture~\cite{nguyen_discontinuous_2014}. Here, since we simulate crack propagation along a weak plane represented as $\Gamma_f$, the above expression is further modified to allow a jump in displacements along $\Gamma_f$. For fracture along an internal interface $\Gamma_f$, the modified expression is given as 
\begin{equation}\label{eq:modified-eq-motion}
\begin{split}
\int_{\Omega} \delta \infinitesimalstrain_{ij}\pkII{ij} \dV{\Omega} &~ - \int_{\Gamma/\Gamma_f} \left( \jump{v}_{i} n^{+}_{j} \right)\avg{\pkII{ij}} \dV{\Gamma}~ -  \\\int_{\Gamma/\Gamma_f} &\left( \jump{u}_{i} n^{+}_{j}\right)  \avg{\pkII{ij}(v)} \dV{\Gamma}~ + \vartheta\int_{\Gamma/\Gamma_f}  \jump{v}_{i}\jump{u}_{i}\dV{\Gamma}~ + \\ &\beta\left\{\int_{\Gamma_f} T_i(\jump{u})\jump{v}_i d\Gamma\right\} ~+ \\ &(1-\beta)\Bigg\{ - \int_{\Gamma_f} ( \jump{v}_{i} n^{+}_{j} )\avg{\pkII{ij}} \dV{\Gamma}~ - \\ &\int_{\Gamma_f} ( \jump{u}_{i} n^{+}_{j}) \avg{\pkII{ij}(v)} \dV{\Gamma}~ + \vartheta\int_{\Gamma_f}  \jump{v}_{i}\jump{u}_{i}\dV{\Gamma}   \Bigg \}
\end{split}
\end{equation}
where $\beta = 1$ if the average traction $\avg{\pkII{ij}} n_j$ along the interface $\Gamma_f$ is greater than the critical  stress,  \ie{} $ \avg{ \pkII{ij} } n_j  \geq \sigma_c$, otherwise $\beta = 0$. The displacement jump  $\jump{u}_i$ ($\jump{u}$ represents the crack tip opening displacement $\delta$ in \cref{fig:setup}) along $\Gamma_f$. The traction-separation relation $T_i(\jump{u})$ (note that $T_i(\jump{u})$ represents $\sigma_{yy}(y=0)$ in \cref{fig:setup}) is a linear cohesive law~\cite{camacho_computational_1996}, characterized by a material's critical strength $\sigma_c$ and fracture energy $\Gamma$ and is described as 
\begin{align}
T_i(\jump{u}) &=  \dfrac{\sigma_c}{g(t)}\left(1-\dfrac{g(t)}{2G_c}\sigma_c\right)\jump{u}_i ~ \mathrm{for~} g(t)=g_\mathrm{max} \\
T_i(\jump{u}) &= \dfrac{\sigma_c}{g_\mathrm{max}}\left(1-\dfrac{g_\mathrm{max}}{2G_c}\sigma_c \right)\jump{u}_i ~ \mathrm{for~} g(t) < g_\mathrm{max}
\end{align}
where $g(t)= \sqrt{\jump{u_n(t)}^2+\beta\jump{u_t(t)}^2}$ and  $g_\mathrm{max}=\underset{t'\in[0, t]}{\mathrm{max}}~ g(t')$. Thus, replacing the second term  in \cref{eq:equation-motion}  with the expression \cref{eq:modified-eq-motion} gives the final equation of motion.

\section{LEFM crack-tip equation of motion}

Linear Elastic Fracture Mechanics (LEFM) describes the growth of a crack by an energy balance
\begin{equation}
    \fractureenergy = \energyreleaserate (\crackhalflength,\crackspeed,\remoteload)~,
    \label{eq:energybalance}
\end{equation}
where $\fractureenergy$ is the fracture energy, assumed to be a constant material property, and $\energyreleaserate$ is the energy release rate that depends on crack (half-)length $\crackhalflength$, crack speed $\crackspeed$, and remote load $\remoteload$. Under assumptions of time-invariant loading, the dynamic energy release rate $\energyreleaserate$ can be related to its static equivalent $\staticenergyreleaserate$ via
\begin{equation}
    \energyreleaserate = \universalfunction \,\staticenergyreleaserate(\crackhalflength,\remoteload)~,
    \label{eq:dynamicenergyreleaserate}
\end{equation}
where $\universalfunction$ is a known universal function~\cite{freund_dynamic_1998}, and can be approximated by $\universalfunction \approx 1 - \crackspeed/\Rwavespeed$. 

Using Eqs.~(\ref{eq:energybalance}) and (\ref{eq:dynamicenergyreleaserate}), one can formulate the crack-tip equation of motion as
\begin{equation}
    \crackspeed/\Rwavespeed \approx  1 - \fractureenergy/\staticenergyreleaserate ~.
    \label{eq:generalEOM}
\end{equation}
In a uniformly loaded system, where plane-strain Griffith's nucleation length is $\griffithlength = 2\shearmodulus\fractureenergy  / \pi  (1-\poissonratio) \remoteload^2$, the static energy release rate is given by $\staticenergyreleaserate = \pi (1-\poissonratio) \remoteload^2 \crackhalflength  / 2\shearmodulus$. Therefore, the specific crack-tip equation of motion for a uniformly loaded system can be written as:
\begin{equation}
    \crackspeed/\Rwavespeed \approx 1 - \frac{\griffithlength}{\crackhalflength} ~.
\end{equation}

\section{Effect of Fracture Energy}
The fracture energy value does not affect the crack dynamics (normalized by Griffith's length) both in linear materials, as expected from LEFM, and GNL materials (see Fig.~\ref{fig:crack-dynamics-different-fracture-energy}). 

\begin{figure}[tb]
\centering
\includegraphics[width=0.98\columnwidth]{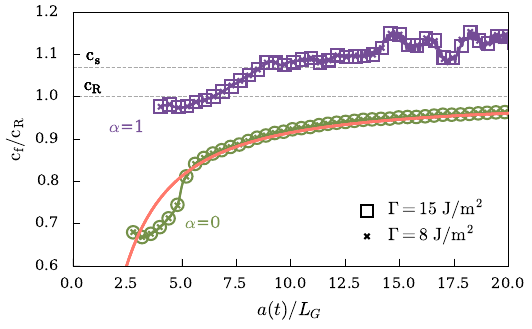}
\caption{Crack-tip dynamics for two different values of fracture energy $\Gamma=15~\mathrm{J/m^2}$ (indicated by hollow square and circle markers) and $\Gamma=8~\mathrm{J/m^2}$ (indicated by cross markers). The crack dynamics is shown in green and purple shades for $\alpha=0$ and $\alpha=1$, respectively. All simulations were conducted at stretch value $\lambda=1.125$. The red solid line is the LEFM crack-tip equation of motion.}
\label{fig:crack-dynamics-different-fracture-energy}
\end{figure}

\section{The near-tip Poynting vector field}
The \textit{instantaneous} rate of energy flow towards the crack tip through an arbitrary contour $\Gamma$ is given by
\begin{equation}\label{eq:poyntng}
    F(\Gamma) = \int_{\Gamma} \underbrace{\Big[\pkI{ji}\dfrac{\partial u_i}{\partial t} + (U+K)\crackspeed \delta_{xj} \Big] }_{P_j}\mathrm{d}\Gamma ~,
\end{equation}
where $\pkI{ji}$ is the first Piola-Kirchhoff stress tensor, $u_i$ is the displacement, $U$ is the strain energy density, $K$ is the kinetic energy density, $\crackspeed$ is the crack speed, and $\delta_{xj}$ is the Kronecker's delta operator, which is 1 only along the crack propagation direction $x$. In \cref{eq:poyntng}, the term inside the integration, denoted as \textit{dynamic} Poynting vector $P_{i}$, represents the direction of the energy flow near the vicinity of the crack tip~\cite{freund_dynamic_1998, buehler_hyperelasticity_2003}. The magnitude of the Poynting vector $\| P\| = \sqrt{P_i}$  is a measure of local energy flow. The quantities in the Pointing vector expression are computed with respect to the reference frame configuration~\cite{freund_dynamic_1998}.

\fi

\end{document}